\documentclass[aps,prb,twocolumn,groupedaddress,floatfix,longbibliography]{revtex4-2}
\usepackage{amsmath,amssymb,graphicx,hyperref}

\usepackage{xcolor}
\usepackage{hyperref}
\hypersetup{
colorlinks,
linkcolor={red!50!black},
citecolor={blue!80!black},
urlcolor={blue!80!black}
}

\renewcommand{\eqref}[1]{Eq.~(\ref{#1})} %I think this is more standard Phys Rev style, but you can comment this out, if you like
\providecommand{\Em}{E_{-}}
\providecommand{\Ep}{E_{+}}
\renewcommand{\vec}[1]{\mathbf{#1}}

\begin{document}

% Use the \preprint command to place your local institutional report
% number in the upper righthand corner of the title page in preprint mode.
% Multiple \preprint commands are allowed.
% Use the 'preprintnumbers' class option to override journal defaults
% to display numbers if necessary
%\preprint{}

%Title of paper
\title{Absorption spectrum of doped highly mismatched alloys}

% repeat the \author .. \affiliation  etc. as needed
% \email, \thanks, \homepage, \altaffiliation all apply to the current
% author. Explanatory text should go in the []'s, actual e-mail
% address or url should go in the {}'s for \email and \homepage.
% Please use the appropriate macro foreach each type of information

% \affiliation command applies to all authors since the last
% \affiliation command. The \affiliation command should follow the
% other information
% \affiliation can be followed by \email, \homepage, \thanks as well.
\author{Hassan Allami}
%\email[]{Your e-mail address}
%\homepage[]{Your web page}
%\thanks{}
%\altaffiliation{}
\affiliation{Department of Physics, University of Ottawa, Ottawa, ON K1N 6N5, Canada}
\author{Jacob J.\ Krich}
\affiliation{Department of Physics, University of Ottawa, Ottawa, ON K1N 6N5, Canada}
\affiliation{School of Electrical Engineering and Computer Science, University of Ottawa, Ottawa, ON K1N 6N5, Canada}

%Collaboration name if desired (requires use of superscriptaddress
%option in \documentclass). \noaffiliation is required (may also be
%used with the \author command).
%\collaboration can be followed by \email, \homepage, \thanks as well.
%\collaboration{}
%\noaffiliation

%\date{\today}

\begin{abstract}
Highly mismatched alloys (HMA's) are a class of semiconductor alloys with large electronegativity differences between the alloying elements.
We predict  the absorption spectrum due to transitions between the split bands of a doped highly mismatched alloy with a conduction band anticrossing. 
We analyze the joint densities of states for both direct and indirect transitions between the split bands. The resulting  spectrum has features that reveal the unusual state distribution that is characteristic of HMAs, hence providing valuable insight into their electronic structure.
In particular, we predict a peak near the absorption edge, which arises due to the suppression of direct transitions at large momenta.
We present analytic forms for the near-absorption-edge and large-energy spectra, showing that they are qualitatively different from those in standard parabolic semiconductors. In particular, as a result of suppressed direct transitions, indirect transitions dominate the spectrum away from the edge of absorption.
\end{abstract}

% insert suggested keywords - APS authors don't need to do this
%\keywords{}

%\maketitle must follow title, authors, abstract, and keywords
\maketitle

\section{Introduction \label{sec:intro}}
Semiconductor alloys where the electronegativity or size of the alloying elements significantly differs are called highly mismatched alloys (HMA's). The hallmark of HMA's, as in the prototypical case of GaAsN,  is a bandgap that changes with alloying in a way that cannot be  explained with a simple bowing parameter \cite{GaAsN-bandgap-drop-92},
a feature that found many applications in making various optoelectronic devices \cite{GaN-led,4th-multi-SC}.
According to the band anticrossing  (BAC) model, the large decrease in the valence-to-conduction band gap with alloying is associated with a second band gap that opens between two split bands $\Em$ and $\Ep$. According to the BAC, these split bands represent the hybridization of localized states from the alloying element with the conduction band of the host semiconductor (See Fig.~\ref{fig:bands_spect}) \cite{bac-original}.
When the localized states hybridize with a single conduction band (CB) of the host, the gap opens by splitting the CB into two split bands, $\Em$ and $\Ep$.
In such HMA's with a CB anticrossing, the splitting can generate a narrow intermediate band, $\Em$, in the original bandgap of the host.
The presence of the narrow band makes HMA's a candidate for implementing intermediate band solar cells \cite{Lopez11,Ahsan12,Welna17,Tanaka17a,Heyman17,Zelazna18,Heyman18},
which have the potential to break the Shockley-Queisser limit on solar cell efficiency \cite{luque-marti-1997}.
Going beyond the BAC, the HMA electronic structure has unusual properties distinct from standard semiconductors \cite{bac-gf}, 
which can manifest in their plasmonic properties when doped \cite{HMA_plasmon}.

Although the absorption spectrum between the two split bands of an HMA is crucial for operating such intermediate band solar cells, there is not much known about it.
While there has been success in doping HMA's so the $\Em$ band is partially full at equilibrium \cite{cl-dope-ZnTeO-19}, the absorption spectra of doped HMA's have not been studied.
The optical absorption of such doped systems is a powerful tool for studying their band structure, which will add to our growing knowledge of these alloys \cite{HMA-special-topic}.

In this work, we predict the absorption spectra of doped highly mismatched alloys with a conduction band anticrossing. We predict the qualitative features of the $\Em$ to $\Ep$ HMA absorption spectrum by calculating the direct and indirect joint densities of states of the system.  We show how these spectra reflect the special electronic structure that arises from the hybridization of localized states with band states, which is distinct in physics and in absorption signatures from absorption in standard semiconductors.
The characteristic predicted signature is a visible peak near the edge of absorption, which is distinguishable from a possible excitonic peak.
We focus on the the case of ZnCdTeO to illustrate these phenomena and their qualitative features. 

\section{absorption spectrum and joint density of states \label{sec:theory}}
While the BAC model \cite{bac-original} is very successful in describing the energy levels in the band structure of HMA's,
it is quiet about how the propagating states are distributed among these energies.
The average Green's function introduced by Wu et al. \cite{bac-gf} captures this special aspect of state distribution in HMA's.
Here, building on their average Green's function, we show how the special state distribution should lead to qualitative features in the absorption spectrum of doped HMA's.

\subsection{HMA's special distribution of states and the absorptivity \label{sec:concepts}}

\begin{figure}[h!]
	\includegraphics[width=\columnwidth]{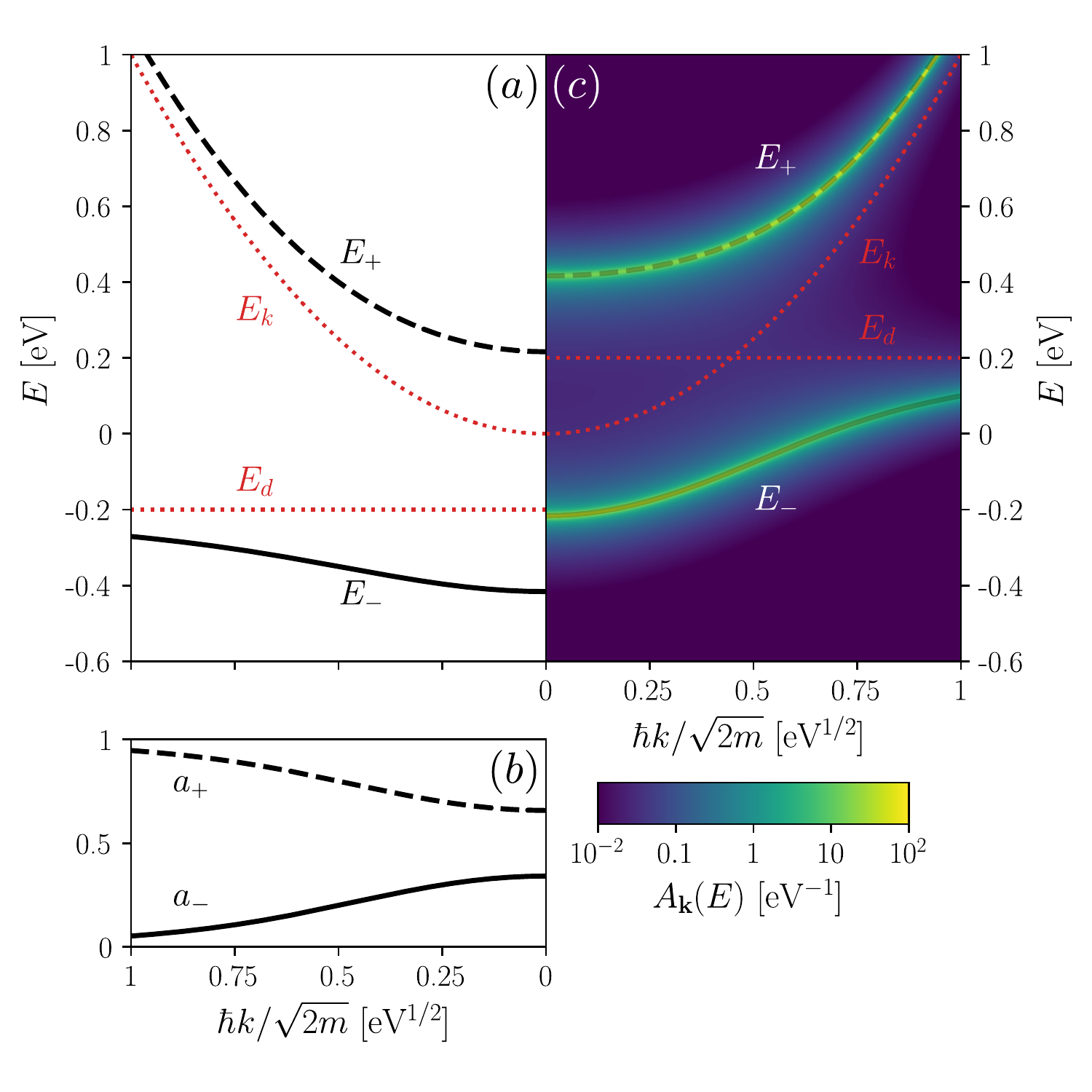}
	\caption{(a) Band anticrossing bands $E_\pm$ according to \eqref{eq:Epm},
	and (b) their corresponding spectral weighting factors $a_\pm$ from \eqref{eq:weights},
	with defect energy $E_d = -0.2 \text{ eV}$ and broadening $\Gamma=0$. 
	(c) Spectral density $A_\vec{k}(E)$ for  $E_d = 0.2 \text{ eV}$ and $\Gamma = 100$~meV
	from  \eqref{eq:A_broken}.
	Note the logarithmic colorscale.
	For reference, $E_\pm$ according to \eqref{eq:Epm} are also plotted in (c).
	In both cases $V = 3$~eV and $x = 1\%$.
	\label{fig:bands_spect}}
\end{figure}

According to the BAC model \cite{bac-original}, large differences in the electronegativities of the mismatched elements compared to the host leads to the formation of localized states with energy $E_d$. These localized states hybridize with the propagating states of the unalloyed host, which have dispersion $E_\vec{k}$, 
resulting in the emergence of two new split bands with dispersions
\begin{equation}
	E_\pm = \frac{1}{2}\left(E_\vec{k} + E_d \pm 
	\sqrt{(E_\vec{k} - E_d)^2 + 4V^2 x}\right),
	\label{eq:Epm}
\end{equation}
where $V$ is the coupling energy between the localized and the propagating states, and $x$ is the alloy fraction of the mismatching element (see Fig.~\ref{fig:bands_spect}).

Although \eqref{eq:Epm} is a great approximation for the energy dispersion of $E_\pm$,
the split bands of HMA's are not just two regular bands arising from a periodic potential of a crystal.
They are both offspring of the hybridization of a single band with a series of randomly distributed localized states.
If we assume that the alloying elements are uniformly distributed, as in the coherent potential approximation \cite{elliott-cpa-review},
we may still use the crystal momentum $\vec{k}$ as a good quantum number in an ensemble-average sense.
The exact hybridized eigenstates with energy $E$ have a non-zero projection on each $\vec{k}$ state.
As a result, at each $\vec{k}$, there is a distribution $A_\vec{k}(E)$, known as the spectral density, that describes the  
projection of all hybridized states with energy $E$ on the $\vec{k}$ states.
As we show in Section~\ref{sec:calc}, the spectral density $A_\vec{k}(E)$ can be broken into two branches, $A_\vec{k}^\pm(E)$, corresponding to $E_\pm$.

Now consider a doped HMA where some excess electrons are occupying the $E_\pm$ bands.
Under illumination with radiation of frequency $\omega$, suppose we want to calculate transition rates of electrons from $\Em$-energy states to $\Ep$-energy states.
In standard semiconductors, the strongest transitions are direct, conserving $\vec{k}$ between initial and final states. 
To find the direct transition amplitudes, at each $\vec{k}$ we must know how many states there are in each band. 
For example, the density of spin up electrons at $\vec{k}$ belonging to the $\Em$ band is $\int  dE A_\vec{k}^-(E)f_E$,
instead of the usual $f_{\Em(\vec{k})}$, where $f_E = (e^{(E-\mu)/T} + 1)^{-1}$ is the Fermi distribution at temperature $T$ and chemical potential $\mu$.

Standard discussion of absorptivity in semiconductors begins from a Fermi's golden rule analysis, generally called
the Kubo-Greenwood formula \cite{marder-absorption}. The absorptivity for photons with energy $E$ is controlled
most importantly by the joint density of states (jDOS) at energy $E$, which describes the number of pairs of states in the 
lower-energy and higher-energy bands that are separated by energy $E$.
Calculating the jDOS, we need to take $A_\vec{k}$ into account, as it carries important information about the special state distribution of HMA's.
For direct absorption processes, the relevant jDOS is
$D_j(E)$, which includes only pairs of states with the same momentum $\vec{k}$. For indirect absorption processes, 
the relevant jDOS is $\rho_j(E)$, which includes all pairs of states separated by energy $E$, regardless of $\vec{k}$.
For both contributions to the optical absorption,
there is a matrix element that must be averaged over all contributing initial and final states. 
The result is that the absorptivity due to direct processes $\alpha_d \propto |M_D|^2 D_j(E)/E$ and 
the absorptivity due to indirect processes $\alpha_i \propto |M_\rho|^2\rho_j(E)/E$,
where $M_D$ and $M_\rho$ are the averaged matrix elements.

We consider both direct momentum-conserving transitions between $\Em$ and $\Ep$ and indirect transitions that do not conserve momentum.
Indirect transitions are typically made possible through phonon exchange,
but in the case of an alloy are also possible without phonons, due to the disorder.
Absorption and emission of phonons that make the indirect transitions possible bring a small shift to the edge of the optical absorption spectrum, which we neglect as it does not affect the qualitative features we describe here.
Neglecting the energy dependence of the matrix elements and the phonon energy shift \cite{marder-absorption}, 
we have
\begin{equation}
	\alpha(E) \propto \frac{D_j(E) + v \rho_j(E)}{E},
	\label{eq:alpha_jdos_relation}
\end{equation}
where $v = |M_{\rho}|^2 / |M_D|^2$ is a volume scale of the system, which we treat as a free parameter.

Calculating the jDOS's for HMA's must account for 
the distribution of electrons at different energies according to $A_\vec{k}^\pm(E)$.
So, we write the direct and indirect jDOS as
\begin{alignat}{2}
	\label{eq:Dj_formal}
	D_j(E) =& \frac{1}{\mathcal{V}}\int &&dE_1dE_2 \sum_{\vec{k}}
	A_\vec{k}^-(E_1)A_\vec{k}^+(E_2)f_{E_1}(1-f_{E_2}) \nonumber\\
	&&&\delta(E_2 - E_1 - E),\\
	\rho_j(E) =& \frac{1}{\mathcal{V}^2}\int && dE_1 dE_2 \sum_\vec{kk'}
	A_\vec{k}^-(E_1)A_\vec{k'}^+(E_2)
	f_{E_1}(1-f_{E_2})\nonumber\\
	&&&\delta(E_2 - E_1 - E),
	\label{eq:rhoj_formal}
\end{alignat}
where $\mathcal{V}$ is the volume of the system.
Comparing the dimension of \eqref{eq:Dj_formal} and \eqref{eq:rhoj_formal} also shows why the ratio $v = |M_{\rho}|^2 / |M_D|^2$ has the dimension of volume.
Note that we neglect the spin degree of freedom --
unless otherwise mentioned --
as it does not affect the shape of $\alpha(E)$.
The presence of  $A_\vec{k}^\pm$ in the jDOS expressions
produces the signatures of the special state distribution of $E_\pm$ in the absorption spectrum of a doped HMA.

\subsection{HMA spectral density and resulting joint density of states \label{sec:calc}}
We now discuss how the special distribution of states in HMA's, given by $A_\vec{k}(E)$, generates special features in $\alpha$ according to \eqref{eq:alpha_jdos_relation}. We begin by deriving $A_\vec{k}(E)$.

Based on Anderson's impurity model \cite{anderson1961}, Wu et al.\ built an average Green's function for electrons in the conduction bands of an HMA \cite{bac-gf}
\begin{equation}
	G(E,\vec{k}) = \left[E - E_{\vec{k}} -
	\frac{V^2 x}{E -E_d + i\Gamma}\right]^{-1},
	\label{eq:the_GF}
\end{equation}
which successfully recovers the spectrum of the BAC model.
The new parameter $\Gamma= \pi\beta V^2\rho_0(E_d)$ determines the broadening of the Green's function's spectral density, where $\rho_0$ is the unperturbed density of propagating states in a unit cell and has dimension of inverse energy.
In what follows we consider a generic case where $E_\vec{k}$ is a parabolic conduction band and all energies are measured from its edge.
Therefore, for instance, $\Gamma \to 0$ for $E_d < 0$, as $\rho_0 = 0$ below the conduction band edge.

While \eqref{eq:the_GF} and the BAC model contain the same energy spectrum in the limit that $\Gamma\to 0$, 
even in that limit the spectral density $A_\vec{k}(E)~=~-{\rm Im}[G(E, \vec{k})]/\pi$ contains information about what fraction of an electron can be
in each of the states. Since there is less than one alloying atom per unit cell of the host crystal, the $\Em$ and $\Ep$ bands can not each
hold as many electrons as the original $E_\vec{k}$ band.

We can divide $A_\vec{k}(E)$ into two pieces, corresponding to $E_\pm$ as
\begin{equation}
	A_\vec{k}(E) = A_\vec{k}^+(E) + A_\vec{k}^-(E)
	=\frac{1}{\pi}\sum_{s=\pm}{\rm Im}\left[\frac{\tilde{a}_s}{E-\tilde{E}_s}\right],
	\label{eq:A_broken}
\end{equation}
where, including the effects of $\Gamma$, the dispersion of \eqref{eq:Epm} is generalized to 
\begin{equation}
	\tilde{E}_\pm = \frac{1}{2}\left(E_{\vec{k}} + \tilde{E}_d  \pm
	\sqrt{(E_{\vec{k}} - \tilde{E}_d)^2 + 4V^2x}\right),
	\label{eq:roots}
\end{equation}
with $\tilde{E}_d = E_d + i \Gamma$, and the generalized weight factors are
\begin{equation}
	\tilde{a}_\pm = \pm \frac{V^2 x}{(\tilde{E}_+ - \tilde{E}_-)(\tilde{E}_\pm - E_{\vec{k}})}.
	\label{eq:a_general}
\end{equation}
A realization of $A_\vec{k}(E)$ with a finite $\Gamma$ is shown in Fig.~\ref{fig:bands_spect}(c). Note that most of the spectral weight is near the BAC bands, but the weight spreads out to nearby energies, as well.

When $\Gamma\to0$, \eqref{eq:roots} reduces to \eqref{eq:Epm}
and the spectral density reduces to two delta functions at $E_\pm$
\begin{equation}
	\lim_{\Gamma\to 0}{A_\vec{k}(E)} = a_+ \delta(E- \Ep) + a_- \delta(E- \Em),
	\label{eq:sharp_A}
\end{equation}
with
\begin{equation}
	a_\pm = \pm \frac{V^2 x}{(E_+-E_-)(E_\pm - E_\vec{k})},
	\label{eq:weights}
\end{equation}
which we derived previously to show how the state distribution in HMA's affects their plasmonic properties \cite{HMA_plasmon}.
The weight factors $a_\pm$ in \eqref{eq:weights} are positive numbers smaller than 1, representing the share of a single $\vec{k}$ state in each of $E_\pm$.
A realization of $a_\pm$ is shown in Fig.~\ref{fig:bands_spect}(b) in a case with $E_d<0$ where $\Gamma\to 0$. 

In the limit  $\Gamma\to0$, we can find analytic forms for Eqs.~(\ref{eq:Dj_formal},\ref{eq:rhoj_formal}). These results are exact for the $E_d<0$ case. While the analytic forms are not exact when $E_d>0$, the analytic results from the $\Gamma\to0$ limit provide useful insight to the behavior of $D_j$ and $\rho_j$ with finite $\Gamma$ as well.
We derive these analytic forms and their implications for finite $\Gamma$ cases in Appendix~\ref{sec:Gamma=0}.
Table~\ref{tab:asymptot} summarizes the qualitative results, showing the scaling of $D_j$ and $\rho_j$ with energy near their respective energy onset and at large energy. 

The most significant prediction  is that the direct $E_-$ to $E_+$ optical transition is suppressed at higher energy,
where the BAC bands alone would predict that  transitions should still be present.
The $E_-$ states at high $k$ have mostly localized character, and as a result the spectral density $a_-$ is small, as is visible in Fig.~\ref{fig:bands_spect}. Since there is little weight in these large-$k$ states, $D_j(E)$ is suppressed, producing a peak in the direct optical absorption spectrum.
In particular, as we show in Appendix~\ref{sec:Gamma=0},
$D_j\sim E^{-3/2}$ for large $E$, as opposed to the usual $\sqrt{E}$ behavior in direct-gap parabolic semiconductors.
Further $\rho_j\sim \sqrt{E}$, as opposed to $E^2$ behavior in indirect parabolic semiconductors.
Note that the jDOS for transitions from an isolated defect to a parabolic band also shows $\sqrt{E}$ behavior, as it mimics the single-particle DOS of the receiving band (see Table~\ref{tab:asymptot}). 

\begin{table}[ht]%[H] add [H] placement to break table across pages
	\caption{Asymptotic behavior of $D_j$ and $\rho_j$ in $\Gamma\to 0$ limit at the absorption edge and for large $E$.
	The two cases of negative and positive $E_d$ are compared with a semiconductor with parabolic bands.
	$E_D$ is the edge of direct transitions, and $E_\rho$ is the edge of indirect transitions.
	The edge behavior of $D_j$ and $\rho_j$ is discussed in Appendix~\ref{sec:Gamma=0}.}
	\label{tab:asymptot}
	\begin{ruledtabular}
		\begin{tabular}{ccccc}
		& \multicolumn{2}{c}{$D_j$} & \multicolumn{2}{c}{$\rho_j$} \\
		 & Edge & Large $E$  & Edge & Large $E$ \\
		\hline
		$E_d < 0$ HMA & $\sqrt{E-E_D}$ & $E^{-3/2}$ & $E-E_\rho$ & $\sqrt{E}$\\
		$E_d > 0$ HMA & $1/\sqrt{E-E_D}$ & $E^{-3/2}$ & $E-E_\rho$ & $\sqrt{E}$\\
		\begin{tabular}{c} parabolic \\ semiconductor\end{tabular} 
			& $\sqrt{E-E_D}$ & $\sqrt{E}$ & $(E-E_\rho)^2$ & $E^2$\\
		\end{tabular}
	\end{ruledtabular}
\end{table}

The presence of the peak in the direct optical absorption is our main prediction,
and according to \eqref{eq:alpha_jdos_relation} should be visible in plots of $E\alpha(E)$. 
Ordinary parabolic semiconductors do not show a similar qualitative feature from their direct band-to-band absorption.
As we show in the Appendix~\ref{sec:Gamma=0},
the width of the peak is controlled by $|E_d|$ when $E_d < 0$ and by $\Gamma$ when $E_d>0$.
Hence this peak also provides a practical way to estimate $\Gamma$, which is otherwise a difficult quantity to measure.

We do not know the ratio of indirect to direct matrix elements $v$, but in general in semiconductors indirect processes are weaker than direct ones, as they need to couple phonons in to the system.
However, since HMA's are random alloys and not crystalline semiconductors, it is possible that momentum conservation between states holds less strongly than in standard semiconductors, which would elevate the indirect $\rho_j$ contribution with respect to the direct $D_j$ contribution to the absorption spectrum. Indirect processes are most important at energies where the direct processes are forbidden, and we predict the form of the $\Em$ to $\Ep$ absorption spectrum assuming the same pattern will hold. 
We predict that since $D_j$ decays as $E^{-3/2}$ for large $E$, while $\rho_j$ grows as $\sqrt{E}$, then if $v$ is large enough the indirect transitions will dominate the large-$E$ part of the $\Em$ to $\Ep$ absorption spectrum.
Ref.~\onlinecite{Heyman17} on the transient absorption spectrum of HMA's invoked indirect transitions to explain their observed high-energy absorption.
Here we provide a firm theoretical basis for indirect transitions dominating direct transitions at higher energy. 
The indirect absorption edge occurs at lower energy than the direct absorption edge, allowing indirect absorption also to dominate for energies below the direct absorption edge. This effect is most noticeable when $E_d<0$, since the direct absorption spectrum is broadened by $\Gamma$ when $E_d>0$. 

All of these features are the direct consequence of the special distribution of states in HMA's, and detecting them in experiments would be a good validation test for the theory.

\section{Experimental Signatures \label{sec:signature}}
We illustrate the predictions for the signatures of $\Em$ to $\Ep$ optical absorption by considering HMA's from the ZnCdTeO family, where the BAC parameters have been estimated \cite{Tanaka_2016,adachi_alloy}. 
Zn$_{1-y}$Cd$_y$Te$_{1-x}$O$_x$ is a II-VI quaternary HMA in which ZnCdTe forms the standard semiconductor and oxygen plays the role of mismatching element.
It has been the subject of extensive studies and used as an HMA of choice in making devices
\cite{TANAKA_13_ZnCdTeO,cl-dope-ZnTeO-19,Tanaka17a,Tanaka_2016,ZnTeO-IBSC,tanaka_2012,Welna_2015,welna_ZnCdTeO_19}.
We  choose ZnCdTeO for our case study because there have been successful attempts in doping the $\Em$ band with chlorine donors \cite{cl-dope-ZnTeO-19}.
Moreover,
controlling the Cd concentration allows for sweeping $E_d$.
BAC parameters for the ternary endpoint alloys are given in Table~\ref{tab:ZnCdTeO}, which shows
that by increasing Cd fraction,
$E_d$ moves from negative to positive.
For the sake of demonstration we show the results for the two ternary endpoints,
ZnTeO and CdTeO, but  
tuning to the Cd fraction allows realization of any $E_d$ between $-0.27$ and $0.38$~eV, 
along with changes in $V$ and $m$, as shown in Table~\ref{tab:ZnCdTeO}.

\begin{table}[t]%[H] add [H] placement to break table across pages
	\caption{Parameters of BAC model for Zn$_{1-y}$Cd$_y$Te$_{1-x}$O$_x$
	\cite{Tanaka_2016,adachi_alloy}, where $m_e$ is the free electron mass.}
	\label{tab:ZnCdTeO}
	\begin{ruledtabular}
		\begin{tabular}{lcc}
		Parameters & $y = 0$ & $y = 1$\\
		\hline
		$E_d$ [eV] & -0.27 & 0.38\\
		$V$ [eV] & 2.8 & 2.2\\
		$m$ [$m_e$] & 0.117 & 0.09\\
		\end{tabular}
	\end{ruledtabular}
\end{table}

\begin{figure}[h!]
	\includegraphics[width=\columnwidth]{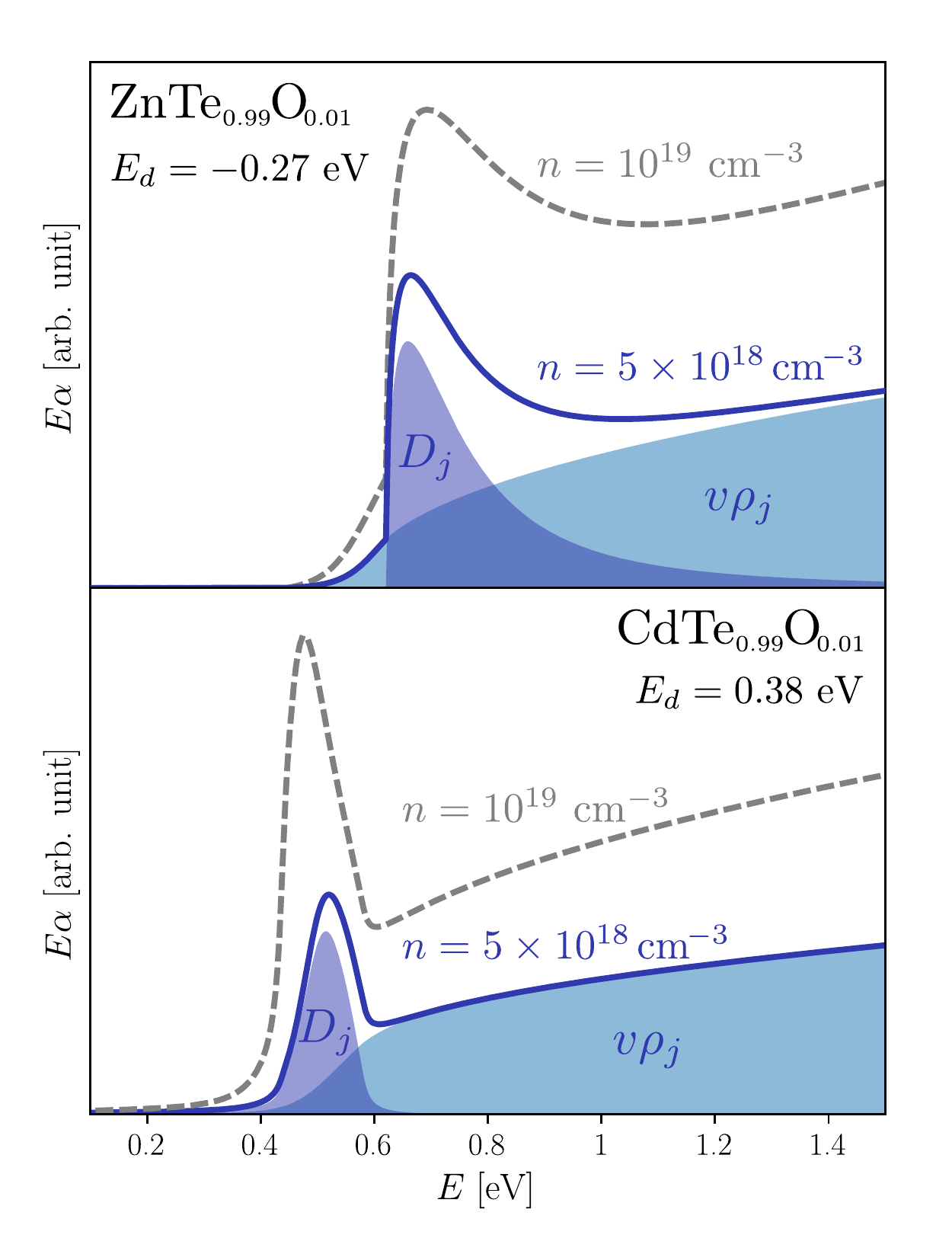}
	\caption{$\Em$ to $\Ep$ optical absorption $E\alpha$ at $T = 300 \text{ K}$ for  ZnCdTeO alloys with two doping levels,
	$n = 5 \times 10^{18} \text{ cm}^{-3}$ (solid blue),
	and $n = 10^{19} \text{ cm}^{-3}$ (dashed grey).
	The separate contributions of direct and indirect transitions also shown for the lower doping level
	(shaded blue),
	according to \eqref{eq:alpha_jdos_relation},
	where $v = 10~\text{nm}^3$ is used for all cases.
	(top)~ZnTe$_{1-x}$O$_x$, with $x=1\%$, and BAC parameters listed in Table~\ref{tab:ZnCdTeO},
	for which $\Gamma \to 0$.
	(bottom)~CdTe$_{1-x}$O$_x$, with $x=1\%$, BAC parameters listed in Table~\ref{tab:ZnCdTeO},
	and $\Gamma = 10 \text{ meV}$ for broadening.
}
	\label{fig:alpha}
\end{figure}

In parabolic band semiconductors, $\alpha(E)$ shows a peak 
due to the factor of $E$ in the denominator of \eqref{eq:alpha_jdos_relation}. 
Instead, we consider  $E\alpha$  to detect the unusual peak of doped HMA's.
While experimentally we generally specify doping level $n$, the absorption depends most directly on the chemical potential appearing in the Fermi function $f_E$. We relate them using
\begin{equation}
	n = \frac{2}{\mathcal{V}}\int dE\sum_\vec{k}A_\vec{k}(E)f_E,
	\label{eq:density}
\end{equation}
where 
the factor of 2 accounts for spin degeneracy.
In Fig.~\ref{fig:alpha} we plot $E\alpha$ at $T = 300$~K for doped ZnTeO (top) and CdTeO (bottom),
according to \eqref{eq:alpha_jdos_relation},
 for two  doping levels. 
For the case with a lower doping level, we also show the contributions of direct and indirect transitions to $E\alpha$ separately (shaded blue).
We use $v = 10~\text{nm}^3$,
chosen strong enough so that $\rho_j$ visibly dominates $D_j$ away from the edge of absorption.
To calculate $D_j$ and $\rho_j$ for CdTeO we used the full form in Eqs.~(\ref{eq:Dj_formal},\ref{eq:rhoj_formal}).
For ZnTeO, since $E_d <0$ and $\Gamma\to 0$, we use
\eqref{eq:Dj_sharp_Edn} to compute $D_j$ and \eqref{eq:rhoj_sharp} to calculate $\rho_j$.

The absorption peak from the direct transitions is clear for both materials in Fig.~\ref{fig:alpha}.
We can see that the CdTeO ($E_d>0$) peak is narrower than the one in ZnTeO ($E_d<0$), which 
we expect as the $E_d>0$ cases have peak widths controlled by $\Gamma$ while the $E_d<0$
have them controlled by $|E_d|$, as discussed in Appendix~\ref{sec:Gamma=0}.
And since  $\Gamma\propto \sqrt{E_d}$ when $E_d>0$,
we expect to see narrower peaks for cases with smaller $|E_d|$.

Figure \ref{fig:alpha} also demonstrates that the indirect transitions dominate at large $E$ in both material systems, due to the decaying large-$E$ tail of $D_j$.
Also, the indirect transitions become available at lower energy
-- as they are not limited by conservation of momentum --
but this effect is only visible for the $E_d < 0$ case where $\Gamma\to 0$ doesn't smear out the edge of absorption.
Since $\Gamma$ being zero in the $E_d < 0$ case gives sharper features to $\alpha$,
we also expect the edge of $D_j$ to be visible as a kink in the absorption in these cases, as shown for ZnTeO.

To show the effects of increased filling of the $\Em$ band,
we include two doping levels in Fig.~\ref{fig:alpha},
$n = 5 \times 10^{18} \text{ cm}^{-3}$ (solid blue),
and $n = 10^{19} \text{ cm}^{-3}$ (dashed grey), for both ZnTeO and CdTeO.
It is clear that the special features of $E\alpha$ --
the absorption peak and the domination of indirect transitions at large $E$ --
are present at both doping levels.
It is possible that excitonic peaks, which generally occur at energies just below the jDOS onset \cite{reynolds2012excitons}, could be confused for the peak that we predict. The evolution of the peak as a result of increasing doping also provides a way to distinguish it from a possible excitonic peak.
In a doped HMA, both screening from carriers in the $\Em$ band and the disordered potential should suppress excitons \cite{doped-exciton}, so we expect the possible excitonic peak to diminish at higher doping.
As Fig.~\ref{fig:alpha} shows, the peak that we expect to be visible in $E\alpha$ naturally grows stronger at higher doping, making it distinguishable from an excitonic one.

It is worth noting that although according to Ref.~\onlinecite{cl-dope-ZnTeO-19}
achieving chlorine concentration up to and even higher than
$10^{20}\text{ cm}^{-3}$ is possible,
it is not clear what portion of the dopants are electrically active.
Especially as our model with weighted bands shows,
$\Em$ has a relatively small maximum capacity for carrying excess electrons.
For instance, for ZnTe$_{1-x}$O$_x$ and CdTe$_{1-x}$O$_x$ with $x=1\%$
used for plotting Fig.~\ref{fig:alpha}, the maximum capacity of $\Em$ is about
$5\times 10^{19}\text{ cm}^{-3}$ and $6\times~10^{19}\text{ cm}^{-3}$
respectively. Electrically active doping above those levels will populate the $\Ep$ band, with changes to the 
$\Em$ to $\Ep$ absorption spectrum, similar to the Moss-Burstein shift \cite{Burstein_1954,Moss_1954}.

In the ZnCdTeO system, the typical frequency range for $\Em$ to $\Ep$ absorption 
is around 0.1~--~1~eV, or  a wavelength of 1~--~10~microns. These spectra can be 
observed with FTIR or other techniques appropriate for these long wavelengths.
We look forward to seeing the results of such experiments, to see if they confirm our theoretical predictions.
We believe experimental results for the absorption spectrum of doped HMA's will be instrumental in understanding their interesting electronic structure better.

\begin{acknowledgments}
We acknowledge funding from the NSERC CREATE TOP-SET program, Award Number 497981.
\end{acknowledgments}

\appendix

\begin{widetext}
\section{Analytic forms and asymptotics of  $D_j$ and $\rho_j$ \label{sec:Gamma=0}}

\subsection{Direct joint density of states \label{sec:Dj}}
Using the sharp $A_\vec{k}$ in \eqref{eq:sharp_A},
the $E_1$ and $E_2$ integrals in \eqref{eq:Dj_formal} become trivial.
Furthermore, since $E_\vec{k} = \hbar^2k^2/2m$ is isotropic we can integrate the angles out to obtain a 1-dimensional integral
\begin{equation}
	D_j(E) = \frac{1}{\mathcal{V}}\sum_{\vec{k}}a_-a_+
	f_{E_-}(1-f_{E_+})\delta(E_+ - E_- - E) =
	\frac{V^2x}{2\pi^2}\int_0^\infty \frac{f_{\Em}(1-f_{\Ep})}{(\Ep - \Em)^2}\delta(E_+ - E_- - E)k^2dk.
	\label{eq:Dj_sharp}
\end{equation}
Note that the factor of $a_-a_+$ originates from the nontrivial spectral density of the HMA, 
and this extra factor produces the $(\Ep - \Em)^2$ in the denominator of \eqref{eq:Dj_sharp},
which makes $D_j$ decay for large $E$,
instead of following the standard $\sqrt{E}$ behavior in parabolic semiconductors. 

When $E_d \leq 0$,
$\Ep - \Em$ has a minimum at $k=0$ with the value $\sqrt{E_d^2 + 4V^2x}$.
Therefore, the argument of the delta function in \eqref{eq:Dj_sharp} has a zero if
$E \geq \sqrt{E_d^2 + 4V^2x}$,
determining the edge of $D_j$.
Carrying out the delta function integral is straightforward and gives
\begin{equation}
	D_j(E) = \left(\frac{2m}{\hbar^2}\right)^{3/2}
	\frac{V^2x \sqrt{E_d + \sqrt{E^2 - 4V^2 x}}}{4\pi^2 E \sqrt{E^2 - 4V^2 x}}
	f_{\hat{E}_-}
	\left(1 - f_{\hat{E}_+}\right)
	\Theta\left(E- \sqrt{E_d^2 + 4V^2 x}\right),
	\label{eq:Dj_sharp_Edn}
\end{equation}
where $\hat{E}_\pm = E_d + \frac{1}{2}(\sqrt{E^2 - 4V^2x} \pm E)$,
and $\Theta$ is the unit step function.
One can see that the fraction in \eqref{eq:Dj_sharp_Edn} decays as $E^{-3/2}$ for large $E$ (see Table~\ref{tab:asymptot}).
Since $D_j$ in \eqref{eq:Dj_sharp_Edn} is positive, continuous, and zero at the edge of direct transitions $E_D = \sqrt{E_d^2 + 4V^2x}$,
it must have a peak. The width of that peak is determined by $|E_d|$.
In this case, $D_j(E)$ rises as $\sqrt{E-E_D}$, the same as the direct joint density in parabolic semiconductors (see Table~\ref{tab:asymptot}).
The peak is visible in the top panel of Fig.~\ref{fig:alpha},
which is plotted for an HMA with $E_d < 0$.

When $E_d>0$, $\Ep - \Em$ has a minimum at $k = \sqrt{2mE_d}/\hbar$ with value $2V\sqrt{x}$,
which leads to a  van Hove singularity at $E_D = 2V\sqrt{x}$ with the same form as the van Hove singularities in 1D materials with parabolic bands. That is, for $E\geq E_D$, 
$D_j(E)\sim 1/\sqrt{E-E_D}$, which is divergent at the onset. This 1D-like form arises from the finite $k^2$ volume element when $\Ep - \Em$ reaches its minimum value in \eqref{eq:Dj_sharp}.
The van Hove singularity arises only when $A_\vec{k}$ is sharp as in \eqref{eq:sharp_A}, in the limit of $\Gamma\to 0$ (see Table~\ref{tab:asymptot}).
But when $E_d>0$, $\Gamma$ is non-zero and broadens $A_\vec{k}^\pm$.
Finite $\Gamma$ turns the divergence of $D_j$ into a peak with a width determined by $\Gamma$.
The peak is visible in the bottom panel of Fig.~\ref{fig:alpha},
which shows an HMA with $E_d > 0$.

The other consequence of $\Ep - \Em$ having a minimum at finite $k$ is that the argument of the delta function in \eqref{eq:Dj_sharp} has two zeros when $2V\sqrt{x} < E < \sqrt{E_d^2 + 4V^2x}$,
which leads to two separate contributions to the integral.
However, for large $E$, where the finite $\Gamma$ effect is also unimportant, there is only one contribution to the integral in \eqref{eq:Dj_sharp}. 
Therefore, the large $E$ behavior of $D_j$ is the same for both positive and negative $E_d$ (see Table~\ref{tab:asymptot}).

\subsection{Indirect joint density of states \label{sec:rhoj}}
To evaluate the non-k-conserving density of states in \eqref{eq:rhoj_formal}, it is helpful to first find the single-particle density of states in each band $\rho_\pm(E)$ by performing $\vec{k,k'}$ integrals.
In the sharp case, where $A_\vec{k}$ is given by \eqref{eq:sharp_A},
we can separately define the densities in each band
\begin{equation}
	\rho_\pm(E) = \frac{1}{\mathcal{V}}\sum_\vec{k}a_\pm \delta(E- E_\pm) = 
	\frac{1}{4\pi^2}\left(\frac{2m}{\hbar^2}\right)^{3/2}
		\begin{cases}
		\sqrt{E - \frac{V^2x}{E-E_d}} & \text{for } E \text{ in } E_\pm \\
		0 & \text{otherwise,}
		\end{cases}
	\label{eq:DOS}
\end{equation}
where we used \eqref{eq:weights} for $a_\pm$.
Integrating $\vec{k,k'}$ out in \eqref{eq:rhoj_formal}
we obtain
\begin{equation}
	\rho_j(E)
	= \int dE_1dE_2\rho_-(E_1)\rho_+(E_2)f_{E_1}(1-f_{E_2})\delta(E_2 - E_1 - E)
	= \int dE_1\rho_-(E_1)\rho_+(E_1 + E)f_{E_1}(1-f_{E_1 + E}).
	\label{eq:rhoj_sharp}
\end{equation}
The edge of $\rho_j$ is at the minimum gap between the $\Ep$ and $\Em$ bands, given by
$E_\rho = \frac{1}{2}(\sqrt{E_d^2 + 4V^2x} - E_d)$,
and it rises linearly (see Table~\ref{tab:asymptot}).
In the limit of large $E$, the $E$-dependence in \eqref{eq:rhoj_sharp} comes from $\rho_+$,
which goes as $\sqrt{E}$, as can be seen from \eqref{eq:DOS}.
Therefore, $\rho_j$ shows $\sqrt{E}$ behavior for large $E$,
which is different from the standard $E^2$ behavior in semiconductors with parabolic bands (see Table~\ref{tab:asymptot}).
The more important implication of this result is that since $D_j$ decays as $E^{-3/2}$,
the indirect transitions can dominate the large-$E$ part of the $\Em$ to $\Ep$ absorption spectrum,
where the direct transitions are suppressed.
Fig.~\ref{fig:alpha} shows this effect away from the edge of absorption.

Lastly, we note that non-zero $\Gamma$ smears out the edge of $\rho_j$ and  has no significant effect on its large-$E$ behavior.

\end{widetext}

\bibliography{references}

\end{document}